\documentstyle[epsfig]{aprim10}
\input{epsf}

\newif\ifAMStwofonts

\ifoldfss
  \ifCUPmtlplainloaded \else
    \NewTextAlphabet{textbfit} {cmbxti10} {}
    \NewTextAlphabet{textbfss} {cmssbx10} {}
    \NewMathAlphabet{mathbfit} {cmbxti10} {} 
    \NewMathAlphabet{mathbfss} {cmssbx10} {} 
  \fi
  \ifAMStwofonts
    \ifCUPmtlplainloaded \else
      \NewSymbolFont{upmath} {eurm10}
      \NewSymbolFont{AMSa} {msam10}
      \NewMathSymbol{\upi}     {0}{upmath}{19}
      \NewMathSymbol{\umu}     {0}{upmath}{16}
      \NewMathSymbol{\upartial}{0}{upmath}{40}
      \NewMathSymbol{\leqslant}{3}{AMSa}{36}
      \NewMathSymbol{\geqslant}{3}{AMSa}{3E}

      \let\leq=\leqslant 
       
    \fi
  \fi
\fi 

\ifnfssone
  \newmathalphabet{\mathit}
  \addtoversion{normal}{\mathit}{cmr}{m}{it}
  \addtoversion{bold}{\mathit}{cmr}{bx}{it}
  \newmathalphabet{\mathbfit} 
  \addtoversion{normal}{\mathbfit}{cmr}{bx}{it}
  \addtoversion{bold}{\mathbfit}{cmr}{bx}{it}
  \newmathalphabet{\mathbfss} 
  \addtoversion{normal}{\mathbfss}{cmss}{bx}{n}
  \addtoversion{bold}{\mathbfss}{cmss}{bx}{n}
  \ifAMStwofonts
    \ifCUPmtlplainloaded \else
      \UseAMStwoboldmath
      \makeatletter
      \new@mathgroup\upmath@group
      \define@mathgroup\mv@normal\upmath@group{eur}{m}{n}
      \define@mathgroup\mv@bold\upmath@group{eur}{b}{n}
      \edef\UPM{\hexnumber\upmath@group}
      \new@mathgroup\amsa@group
      \define@mathgroup\mv@normal\amsa@group{msa}{m}{n}
      \define@mathgroup\mv@bold\amsa@group{msa}{m}{n}
      \edef\AMSa{\hexnumber\amsa@group}
      \makeatother
      \mathchardef\upi="0\UPM19
      \mathchardef\umu="0\UPM16
      \mathchardef\upartial="0\UPM40
      \mathchardef\leqslant="3\AMSa36
      \mathchardef\geqslant="3\AMSa3E

      \let\leq=\leqslant 

    \fi
  \fi
\fi 

\ifnfsstwo
  \DeclareMathAlphabet{\mathbfit}{OT1}{cmr}{bx}{it}
  \SetMathAlphabet\mathbfit{bold}{OT1}{cmr}{bx}{it}
  \DeclareMathAlphabet{\mathbfss}{OT1}{cmss}{bx}{n}
  \SetMathAlphabet\mathbfss{bold}{OT1}{cmss}{bx}{n}
  \ifAMStwofonts
    \ifCUPmtlplainloaded \else
      \DeclareSymbolFont{UPM}{U}{eur}{m}{n}
      \SetSymbolFont{UPM}{bold}{U}{eur}{b}{n}
      \DeclareSymbolFont{AMSa}{U}{msa}{m}{n}
      \DeclareMathSymbol{\upi}{0}{UPM}{"19}
      \DeclareMathSymbol{\umu}{0}{UPM}{"16}
      \DeclareMathSymbol{\upartial}{0}{UPM}{"40}
      \DeclareMathSymbol{\leqslant}{3}{AMSa}{"36}
      \DeclareMathSymbol{\geqslant}{3}{AMSa}{"3E}

      \let\leq=\leqslant 

    \fi
  \fi
\fi 

\ifCUPmtlplainloaded \else
  \ifAMStwofonts \else 
    \def\upi{\pi}
    \def\umu{\mu}
    \def\upartial{\partial}
  \fi
\fi

\title[The Apparent ``Angular Size~--~Frequency'' Dependence in $\sim$3000 Parsec-scale Extragalactic Jets]
{The Apparent ``Angular Size~--~Frequency'' Dependence in
$\sim$3000 Parsec-scale Extragalactic Jets }

\author[Yang et al.]
       {J.~Yang$^{1,2}$, L.I. Gurvits$^2$, S. Frey$^{3,4}$, and A.P. Lobanov$^5$\\
        $^1$Shanghai Astronomical Observatory, Chinese Academy of Sciences, 80 Nandan Road, 200030 Shanghai, P.R. China\\
        $^2$Joint Institute for VLBI in Europe, PO Box 2, 7990 AA Dwingeloo, The Netherlands\\
        $^3$F\"OMI Satellite Geodetic Observatory, PO Box 585, 1592 Budapest, Hungary\\
        $^4$MTA Research Group for Physical Geodesy and Geodynamics, PO Box 91, 1521 Budapest, Hungary\\
        $^5$Max-Planck-Institut f\"ur Radioastronomie, Auf dem H\"ugel 69, 53121 Bonn, Germany}
\date{}

\pagerange{\pageref{firstpage}--\pageref{lastpage}}
\pubyear{2008}

\begin{document}

\maketitle

\label{firstpage}

\begin{abstract}
The upper envelope of the amplitude of the VLBI visibility
function usually represents the most compact structural pattern of
extragalactic radio sources, in particular, the ``core-jet''
morphologies. By fitting the envelope to a circular Gaussian model
in $\sim$3000 parsec-scale core-jet structures, we find that the
apparent angular size shows significant power-law dependence on
the observing frequency (power index $n=-0.95\pm0.37$). The
dependence is likely to result from synchrotron self-absorption in
the inhomogeneous jet and not the free-free absorption ($n=-2.5$),
nor the simple scatter broadening ($n\leq-2$).

\end{abstract}


\section{Introduction}

Radio-loud active galactic nuclei (AGNs) are a class of
high-luminosity objects in the Universe. Their pc-scale jets can
be well imaged by VLBI observations. The ``angular
size~--~frequency'' (``$\theta$~--~$\nu$'') relation can be used
for calibrate the apparent ``angular size~--~redshift''
(``$\theta$~--~$z$'') relation in which the radio jets are taken
as a ``standard rod'' applicable for testing different
cosmological models \cite{gurv03}. The dependence of the size on
frequency can provide important constraints for studying the
internal physical properties \cite{loba98} and the evolution of
jets.

An extragalactic radio source generally has a power-law total flux
density spectrum resulting from the synchrotron radiation.
However, when imaged with VLBI at a milliarcsecond resolution, the
jet emission often exhibits more complicated spectral shapes,
affected by opacity in the emitting material itself and in the
surrounding medium. The intervening media along the line of sight
scatter/absorb the emission and distort the image, in particular
at low ($<1$~GHz) frequencies. Thus, the apparent angular size
usually depends on the observing frequency. The dependence can be
written phenomenologically as a power-law function
$\theta\propto\nu^n$ in most cases. For instance, $n=-2$ for the
compact radio source at the center of our Galaxy (Sgr~A$^*$) as a
result of scatter broadening \cite{shen05}.

\begin{table}
\centering \caption{Summary of the used data set.}\label{tab1}

\begin{tabular}{cccrc}
  \hline \hline
  Band & $\nu_{\rm{obs}}$ & $R$ & $N_{\rm{source}}$ & Surveys\\
     &(GHz)&  (mas)  &      &          \\
  \hline
  S  & 2.3 & 3.2  & 3115 & RRFID$^a$ and VCS$^b$   \\
  C  & 5.0 & 1.4  & 1406 & VIBApls$^c$ and VIPS$^d$ \\
  X  & 8.4 & 0.85 & 3004 & RRFID and VCS   \\
  Ku & 15  & 0.47 & 299  & VLBA 2-cm Survey \\
  Ka & 24  & 0.32 & 264  & RRFID  \\
  Q  & 43  & 0.17 & 124  & RRFID \\
  \hline
\end{tabular}
 \vspace{-0.5em}
\begin{flushleft}
{\tiny Columns: (1) Observing band; (2) Observing frequency; (3)
Angular resolution of the VLBA at this band; (4) Number of sources
analysed; (5) Major VLBI surveys from which the
data are obtained.\\
$^a$~RRFID: Radio Reference Frame Image Database \cite{fey00}. \\
$^b$~VCS: VLBA Calibrator Surveys \cite{petr06}. \\
$^c$~VLBApls: VSOP Pre-lunch Survey \cite{foma00}.\\
$^d$~VIPS: VLBA Imaging and Polarimetry Survey
\cite{helm07}.\\}\end{flushleft}
\end{table}

\section{The Characteristic Angular Size}
\begin{figure}
  \includegraphics[clip, width=0.48\textwidth]{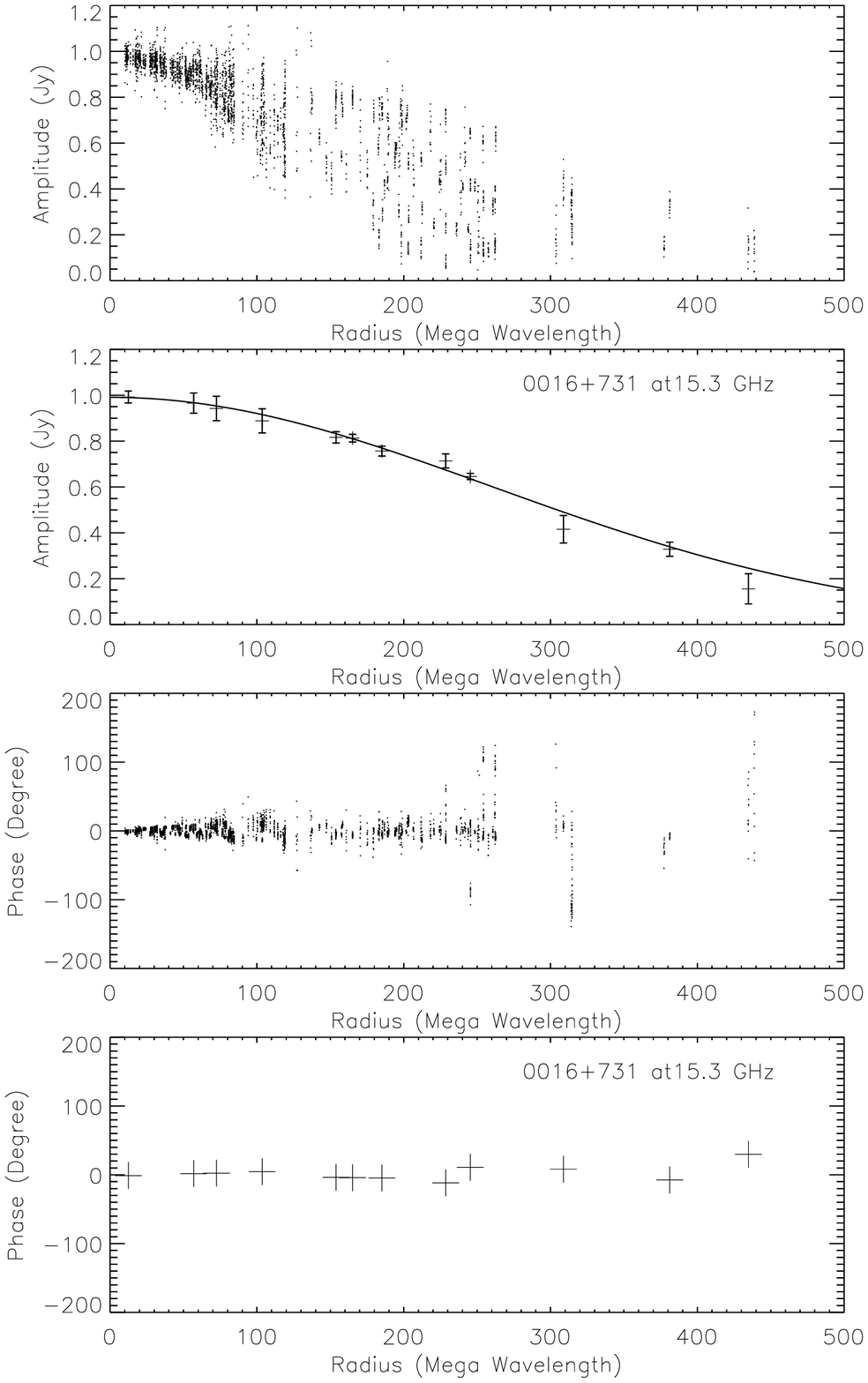}\\
  \caption{An example for the definition of the characteristic
angular size. The top panel shows the original visibility data.
The bottom panel shows the extracted upper envelope points and the
fitting Gaussian curve.}\label{fig1}
\end{figure}

The characteristic angular size is a parameter describing the
width of the intensity distribution of jet emission. Generally, it
can be defined either in the image plane or in the $u$-$v$
(Fourier) plane. For instance, a measure of the characteristic
size can be provided by the distance between the brightest
component and the most distant one with the brightness at least
2\,\% of the peak \cite{kell93} in the image plane. It is also
feasible to obtain the estimated angular size by fitting a
circular Gaussian model to the visibility data \cite{gurv94} in
the $u$-$v$ plane. Although the two ways are equivalent in
principle, the definitions in the $u$-$v$ plane are more direct
than those in the image plane, and free from sampling effect. If
the visibility data have enough sensitivity, the
``super-resolution'' can be obtained for the barely resolved
sources using a certain model \cite{loba05}. Here we define the
characteristic angular size $\theta$ by fitting the upper envelope
of the visibility amplitudes to a circular Gaussian model
\cite{pear99}:
\begin{equation}\label{eq1}
    \Gamma (\rho) = \exp \left[ \frac{- (\pi
    \theta \rho)^2}{4 \ln 2} \right]
\end{equation}
where $\theta$ is the full width to half-maximum intensity (FWHM)
in the image plane in radian, $\rho$ is $u$-$v$ radius in
wavelength and $\Gamma (\rho)$ is the normalized amplitude. The
characteristic angular size represents the most compact and the
brightest part of the source. Fig.~\ref{fig1} gives an example for
the definition of the characteristic angular size.

\section{Data Collection}
Table~\ref{tab1} gives a brief summary of our data set of $~$4000
objects compiled from several large VLBI surveys \cite{frey06}.
Some sources have multi-frequency and/or multi-epoch observations.
All the visibility data have been calibrated, self-calibrated, and
saved as standard $u$-$v$ FITS files. We developed several
programs to automatically calculate the characteristic angular
size according to our definition using the IDL Astronomy User¡®s
Library. We extracted the upper envelope points from the original
visibility data, fitted them by a circular Gaussian model, and
finally fitted the ``$\theta$~--~$\nu$'' data to a power-law
function to determine the power index $n$ for the multi-frequency
observations.

\section{Results}

\begin{figure}
 \centerline{{\epsfxsize=0.48\textwidth\epsffile{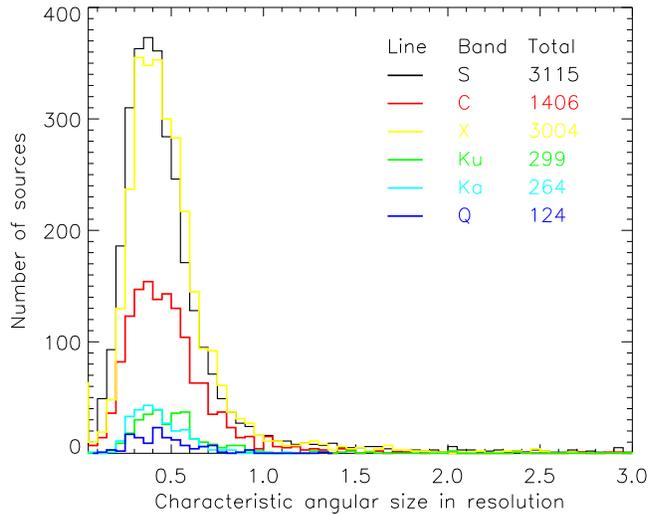}}}
 \caption[]{The statistical distributions of the characteristic
angular sizes. Different colors represent different bands. Note
that the angular size is expressed in units of the resolution of
the VLBA array at each band.}\label{fig2}
\end{figure}

\begin{figure} 
 \centerline{{\epsfxsize=0.48\textwidth\epsffile{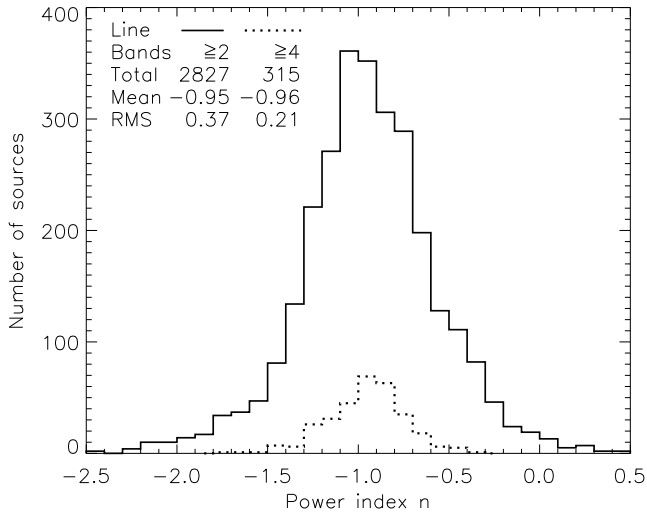}}}
 \caption[]{The statistical distribution of the power index $n$. The solid
line represents the result from all the sources with at least
two-frequency observations. The dotted line represents the result
from all the sources with at least four-frequency observations.}
\label{fig3}
\end{figure}

Fig.~\ref{fig2} shows the statistical distributions of the
characteristic angular sizes. The histograms show a similar shape
peaking at the position of $\sim$0.4$R$, where $R$ is the
resolution of the VLBA array at each band and listed in Column 3,
Table~\ref{tab1}. The Ku-band distribution is consistent with the
previous results in the MOJAVE sample by Kovalev et al.
\shortcite{kova05}. They fitted the core by an elliptical Gaussian
model in Difmap. Fig.~\ref{fig3} shows the distribution of the
power index $n$, which has a mean value of $-0.95\pm0.37$ for 2827
sources with at least two-frequency observations and
$-0.96\pm0.21$ for 315 sources with at least four-frequency
observations. The dependence agrees well with the results from
fitting the core by an elliptical Gaussian model in 167 objects
\cite{jian01} . Note that we assume that the power index $n$ is
independent of redshift within the studied range of redshift
($0<z<5$).

\section{Discussion}
The upper envelope statistically gives the size of the brightest
and most compact component. Among all the components in a pc-scale
jet, the core/jet base usually has a flatter spectrum and
dominates the total flux density. Thus, the upper envelope gives
the size of the radio core in most cases.

The radio core generally has a flat spectrum and is located in the
region where the optical depth $\tau=1$. K\"onigl (1981)
demonstrated that, due to the synchrotron self-absorption, the
size of the core $r_{\rm{core}}$ varies with the frequency $\nu$
as: $r_{\rm{core}}\propto\nu^{-1/k_{\rm{r}}}$, where $k_{\rm{r}}$
depends on the shape of the electron energy spectrum and on the
distribution of the magnetic field and particle density in an
inhomogeneous jet, $k_{\rm{r}}=1$ in the case of the equipartition
between jet particle and magnetic field energy densities
\cite{loba98}. Our results agree well with this prediction.

Free-free absorption by the plasma covering the jet also results
in an optically thick core. However, free-free absorption is only
found in a few sources. Lobanov \shortcite{loba98} suggested that
the shift of core position is $r_{\rm{core}}\propto\nu^{-2.5}$ for
a spherical distribution of free-free absorption plasma
$N_{\rm{e}}\propto{r}^{-3}$, where $r$ is the distance from the
center and $N_{\rm{e}}$ is the electron number density. So, the
free-free absorption is not the right mechanism to explain the
statistical relation presented here. It might however happen in
some sources. As far as the simple diffractive scattering models,
e.g. the Gaussian screen model or the power-law model with the
Kolmogorov spectrum, are concerned, they result in a dependence
with $n\leq-2$ \cite{tomp86}, which is fairly far from the peak
($n\sim-0.9$) of the obtained distribution. Thus, simple scatter
broadening cannot explain the dependence.

\section{Conclusions}
Based on the above discussion, we can draw the following
conclusions.
\begin{enumerate}
\item The upper envelope of visibility amplitudes is a
statistically meaningful characteristic angular size for the radio
``core''.

\item The apparent ``angular size~--~frequency'' dependence can be
statistically described by $\theta\propto\nu^{-0.95\pm0.37}$ in
$\sim$3000 pc-scale jets.

\item The observed frequency dependence can be explained as a
result of synchrotron self-absorption in the inhomogeneous jet.
The simple scatter broadening and free-free absorption can be
ruled out as dominant mechanisms responsible for the dependence.
\end{enumerate}

\section{Acknowledgment}
J.~Yang is grateful to the KNAW-CAS grant 07DP010. S.~Frey
acknowledges the OTKA~K72515 grant. We made use of the collection
of calibrated VLBI $u$-$v$ data maintained by L. Petrov at NASA
Goddard Space Flight Center (GSFC). This research has made use of
NASA's Astrophysics Data System, NASA/IPAC Extragalactic Database
(NED) and the United States Naval Observatory (USNO) Radio
Reference Frame Image Database (RRFID). We thank A.L. Fey for
providing the RRFID data. The National Radio Astronomy Observatory
is a facility of the National Science Foundation operated under
cooperative agreement by associated Universities, Inc.

\label{lastpage}

\clearpage

\end{document}